\documentclass[12pt]{article}

\newcommand{\eref}[1]{eq.\,(\ref{e.#1})} 
\newcommand{\erefn}[1]{(\ref{e.#1})}

\newcommand{\cref}[1]{Chapter \ref{c.#1}}

\def\simlt{\stackrel{<}{{}_\sim}}
\def\simgt{\stackrel{>}{{}_\sim}}   
\def\beq{\begin{equation}} 
\def\eeq{\end{equation}} 
\def\bea{\begin{eqnarray}}  
\def\eea{\end{eqnarray}}  
\def\ba{\begin{array}}  
\def\ea{\end{array}}   
\def\bi{\begin{itemize}}  
\def\ei{\end{itemize}}  
\def\be{\begin{enumerate}}  
\def\ee{\end{enumerate}}  
\def\beq{\begin{equation}}  
\def\eeq{\end{equation}}  
\def\bc{\begin{center}}
\def\ec{\end{center}}

\def\hc{{\rm H.c.}}

\def\cl{{\mathcal L}}  
   
\def\cm{{\mathcal M}}  
\def\co{{\mathcal O}}

\def\t{\tilde}

\def\ov{\overline}

\def\tev{\,{\rm TeV}}
\def\gev{\,{\rm GeV}}

\begin{document}
\pagestyle{empty}
\begin{flushright}
IFT-2004/20\\
{\tt hep-ph/0407242}\\
{\bf \today}
\end{flushright}
\vspace*{5mm}
\begin{center}

{\large {\bf Electroweak symmetry breaking in supersymmetric models
with heavy scalar superpartners}}\\
\vspace*{1cm}
{\bf Piotr~H.~Chankowski}\footnote{Email:chank@fuw.edu.pl}, 
{\bf Adam Falkowski}\footnote{Email:afalkows@fuw.edu.pl},
{\bf Stefan~Pokorski}\footnote{Email:pokorski@fuw.edu.pl},\\ 
and {\bf Jakub Wagner}\footnote{Email:jwagner@fuw.edu.pl}\\
\vspace{0.3cm}

Institute of Theoretical Physics, Warsaw University, Ho\.za 69, 00-681,
Warsaw, Poland\\

\vspace*{1.7cm}
{\bf Abstract}
\end{center}
\vspace*{5mm}
\noindent
{
We propose a novel mechanism of electroweak symmetry breaking in 
supersymmetric models, as the one recently discussed by Birkedal, 
Chacko and Gaillard, in which the Standard Model Higgs doublet is 
a pseudo-Goldstone boson of some global symmetry. The Higgs mass 
parameter is generated at one loop level by two different, moderately
fine-tuned sources of the global symmetry breaking. The mechanism 
works for scalar superpartner masses of order $10$~TeV, but gauginos 
can be light. The scale at which supersymmetry breaking is mediated 
to the visible sector has to be low, of order $100$~TeV. Fine-tuning 
in the scalar potential is at least two orders of magnitude smaller 
than in the MSSM with similar soft scalar masses. The physical Higgs 
boson mass is (for $\tan\beta\gg1$) in the range $120-135$ GeV. 
}
\vspace*{1.0cm}
\date{\today}


\vspace*{0.2cm}

\vfill\eject
\newpage

\setcounter{page}{1}
\pagestyle{plain}

\section{Introduction}

Explanation of the origin of the Fermi scale is a challenge for physics
beyond the Standard Model (SM)\footnote{A different point of view has
recently been expressed in refs. \cite{dakili,ardi}.} 
An interesting possibility is that the electroweak symmetry breaking is
generated by quantum corrections to the Higgs doublet potential. For such
a mechanism to be under theoretical control, the tree level Higgs mass 
squared parameter $m_H^2$ has to be calculable (at least in principle) 
at some scale $\Lambda_S$ in terms of some more fundamental parameters.
Secondly, the dependence of quantum corrections to the Higgs doublet 
potential on dimensionful parameters of new physics should be moderate. 
Otherwise large cancellations between the tree level parameters and quantum 
corrections would be necessary, rendering such a mechanism doubtful.
Although fine-tuning is difficult to quantify in a precise way, it is 
usually easy to give a rough estimate of its order of magnitude. As a 
reference number it is worth remembering that in the SM cut-off by the 
neutrino see-saw scale of order $10^{13}$ GeV, the Higgs potential has 
to be fine-tuned in 1 part to $10^{20}$. 

Radiative electroweak symmetry breaking, triggered by the large top 
quark Yukawa coupling, has been a succesful prediction of the Minimal 
Supersymmetric Standard Model (MSSM). However, in the MSSM the necessary 
degree of fine-tuning in the Higgs potential grows exponentially with the 
value of the lightest CP-even Higgs boson mass \cite{CHELPO}. The present 
LEP limits on the Higgs boson mass push the stop masses into the range 
500~GeV$-$1~TeV and, in consequence, the necessary fine-tuning in the MSSM 
Higgs potential is estimated to be of order of 1\% \cite{CHELOLPO,FEMAMO}. 
This fact may be taken as somewhat dissapointing for a supersymmetric model 
and it stimulated several authors to look for alternatives to the MSSM 
\cite{CAESHI} and to supersymmetry itself, which could explain the origin 
of the Fermi scale \cite{bahano,arcoge,csgrmu,FAGRPO}. However, no convincing 
idea has emerged yet that would lead to fine-tuning significantly lower than 
${\cal O}(1\%)$ needed in the MSSM with the present limits on the Higgs boson 
mass. 

We believe, therefore, it may be worthwhile to ask a different question. 
After all, the questions about an acceptable degree of fine-tuning and even 
about its definition do not have any sensible quantitative answer at the level
of effective theories, for instance, when the theory of soft supersymmetry
breaking terms is not known \cite{CHELOLPO,KAKI}. More bothersome is the 
quadratic dependence of the fine-tuning in the MSSM on the mass of the scalar 
superparners of the top quark. Although the FCNC effects in the MSSM are 
controlled by the squark masses of the first two families, a light stop is 
not easy (although not impossible) to reconcile with the observed suppression 
of the FCNC effects \cite{ARHAMU}. Indeed, unless the first two sfermion 
families are degenerate in mass, they have to be heavy, with masses at least 
${\cal O}(10$~TeV), and the large splitting with the third one needs some 
explanation by the mechanism of supersymmetry breaking \cite{bidu}. Thus it 
is of some interest to ask if in models in which radiative electroweak 
symmetry breaking occurs one can avoid quadratic dependence of the Fermi 
scale on the stop masses, i.e. if one can significantly rise the scalar 
superpartner masses without jeopardizing naturalness.

An interesting and very simple combination of the idea of the Higgs doublet
as a pseudo-Goldstone boson (revived in the non-supersymmetric Little Higgs 
models \cite{arcoge}, inspired by the so-called deconstruction \cite{HIPOWA}) 
and of supersymmetry has been proposed recently by Birkedal, Chacko and 
Gaillard \cite{BICHGA}. In their model the Higgs doublet is a Goldstone 
boson of a spontaneously broken global $SU(3)$ symmetry. Global $SU(3)$ is 
also explicitly broken by a supersymmetric fermion mass term and by the SM 
$SU(2)_L\times U(1)_Y$ electroweak interactions. In the present paper we 
show that in this model, in certain range of its parameters, an interesting
mechanism of radiative electroweak symmetry breaking can be realized. The 
Higgs doublet mass parameter is then generated at one-loop level by the  two 
moderately fine-tuned sources of the global $SU(3)$ symmetry breaking. The 
mechanism works for scalar superpartner masses of order ${\cal O}(10$~TeV). 
Fine-tuning in the scalar potential is at least two orders of magnitude 
smaller than in the MSSM with similar soft scalar masses, i.e. stays at the 
level of ${\cal }(1\%)$. The physical Higgs boson mass is predicted to be 
(for $\tan\beta\gg1$) in the range $120-135$~GeV, where the main source of 
uncertainty are unknown two-loop effects.  

\section{The model}

In this section we introduce our model which is a slight modificaton of the
one proposed in \cite{BICHGA}. The Higgs and the 3rd generation weak doublet 
superfields are extended to fit the fundamental ($\hat{\cal H}_u$) and 
anti-fundamental ($\hat{\cal H}_d$ and $\hat{\cal Q}$) representations of an 
approximate global $SU(3)$ symmetry:
\begin{eqnarray}
\hat{\cal H}_u=\left(\matrix{\hat H_u\cr\hat S_u}\right),\phantom{aa}
\hat{\cal H}_d^T =\left(\matrix{\hat H_d\cr\hat S_d}\right),\phantom{aa}
\hat{\cal Q}^T = \left(\matrix{\hat Q_3\cr\hat T}\right)~.
\end{eqnarray}
In addition there is a new $SU(3)$ singlet quark supermultiplet $\hat T^c$. 
At some high scale $\Lambda_S$ the $SU(3)$ symmetry 
is respected by the top Yukawa couplings in the superpotential, but is 
explicitly broken by the $\mu_T$ term:
\begin{eqnarray}
w_t= Y_T\hat{\cal Q}\hat{\cal H}_u\hat t^c + \mu_T\hat T^c\hat T  
\label{eqn:sup_wt}
\end{eqnarray}
In order to preserve the symmetry between the up- and down- type sectors 
one can introduce another global $SU(3)^\prime$ symmetry, 
that controls the bottom sector. To this end we extend the quark doublet 
$\hat Q_3$ by the $SU(2)_L$ singlet quark superfield $\hat B$, so that
\begin{eqnarray}
\hat{\cal Q}^\prime =  \left(\matrix{\hat Q_3\cr\hat B}\right)
\end{eqnarray}
is in the fundamental representation of $SU(3)^\prime$ (with respect to 
which $\hat{\cal H}_u$ and $\hat{\cal H}_d$ form the fundamental and 
anti-fundamental representations, respectively). We also introduce a 
corresponding $SU(3)^\prime$ singlet quark superfield $B^c$. The bottom 
sector superpotential then reads
\begin{eqnarray}
w_b =Y_B \hat{\cal H}_d\hat{\cal Q}^\prime\hat b^c  + \mu_B\hat B^c\hat B  
\label{eqn:sup_wb}
\end{eqnarray}
Here $SU(3)^\prime$ is explicitly broken by $\mu_B$. Of course, $w_b$ breaks  
$SU(3)$ while $w_t$ breaks $SU(3)^\prime$. We also assume that at the high 
scale $\Lambda_S$ the $SU(3)$ and  $SU(3)^\prime$ symmetries are respected 
by the soft mass terms and trilinear couplings in the top and bottom sectors, 
respectively. Thus, the most general form of these mass terms (at $\Lambda_S$) 
is: \footnote{The requirement of the $SU(3)$ symmetry of the soft terms
eliminates the otherwise allowed $m^2_{3T}\tilde T\tilde T^c$ and
$m^2_{3B}\tilde B\tilde B^c$ terms.}
\begin{eqnarray}
&&{\cal L}_{tb} = 
-m^2_{{\cal Q}_3}(|\tilde Q_3|^2+|\tilde T|^2+|\tilde B|^2)\nonumber\\
&&\phantom{aaaa} - m^2_{U_3} |\tilde t^c|^2 - m_{T^c}^2 |\tilde T^c|^2
                 - m^2_{D_3} |\tilde b^c|^2 - m_{B^c}^2 |\tilde B^c|^2
\label{eqn:softmasses}\\
&&\phantom{aaaa}
+A_T(\tilde Q_3 H_u\tilde t^c + \tilde T S_u \tilde t^c)
+A_B(H_d \tilde Q_3\tilde b^c + S_d \tilde B \tilde b^c)~.\nonumber
\end{eqnarray}
 Finally, the Higgs doublet $\mu$ term is assumed to respect the $SU(3)$ 
(and in fact, also the $SU(3)^\prime$) symmetry 
\begin{eqnarray}
w_h = \mu ~\hat{\cal H}_d\hat{\cal H}_u~.
\label{eqn:sup_mu}
\end{eqnarray}

As in ref. \cite{BICHGA} the gauge symetry of the MSSM is extended by the 
$U(1)_E$ group, which commutes with the global $SU(3)$ and $SU(3)^\prime$  
symmetries. The $U(1)_E$ gauge coupling $g_E$ is normalized in such a way 
that $\hat{\cal H}_u$, and $\hat{\cal H}_d$ have $U(1)_E$ charges $+1$ and 
$-1$ respectively. The quantum numbers of the relevant fields of the model 
are summarized in Table \ref{t:charges}. Since the electroweak symmetry 
should not be broken by the vevs of the $S_u$ and $S_d$ fields one must 
assume that both these superfields are SM singlets. The SM quantum numbers 
of all the new superfields are uniquely determined by the form of the 
superpotential. Furthermore, the anomaly cancellation requirement constrains 
the $U(1)_E$ charges of the SM fields to be proportional to the hypercharge 
(the other possibility that the additional $U(1)$ is proportional to $B-L$ 
cannot be realized here as the Higgs fields must be charged under $U(1)_E$)
\cite{WE}. With our normalization of the Higgs $U(1)_E$ charges this 
determines all the remaining $U(1)_E$ charges uniquely.   

\begin{table}
$\ba{c|c|c|c|c|c|c|c|c|c|c|c|}
& \hat H_u & \hat S_u & \hat H_d & \hat S_d & \hat Q & \hat T & \hat t^c & 
\hat T^c & \hat B & \hat b^c & \hat B^c  
\\ \hline 
SU(3) & 1 & 1 & 1 & 1 & 3 & 3 & \ov 3 & \ov 3 & 3 & \ov 3 & \ov 3
\\ \hline
SU(2)  & 2 & 2 & \ov 2 & \ov 2 & 2 & 1 & 1 & 1 & 1 & 1 & 1
\\ \hline
U(1)_Y   & 1/2 & 0 & -1/2 & 0 & 1/6 & 2/3 & -2/3 & -2/3 & -1/3 & 1/3 & 1/3
\\ \hline
U(1)_E& 1 & 1 & -1 & -1 & 1/3 & 1/3 & -4/3 & -1/3 & 1/3 & 2/3 & -1/3
\ea
$
\caption{Quantum numbers of the multiplets in the model}
\label{t:charges}
\end{table}

The $SU(3)$ preserving  part of the scalar potential of the Higgs fields reads
\begin{eqnarray}
V_{\rm symm}=m_1^2|{\cal H}_d|^2+m_2^2|{\cal H}_u|^2 
            -(m_3^2{\cal H}_d{\cal H}_u + \hc)
            +{1\over2}g_E^2(|{\cal H}_u|^2 - |{\cal H}_d|^2)^2 ~,
\label{eqn:V_su3symm}
\end{eqnarray}
where $m_1^2 =\mu^2 + m_{{\cal H}_d}^2$, $m_2^2 = \mu^2 +  m_{{\cal H}_u}^2$ 
and the parameters $m_{{\cal H}_d}^2$, $m_{{\cal H}_u}^2$, $m_3^2$  are the 
soft breaking mass terms. As we discuss in more detail in Sec. \ref{sec:4}, 
spontaneous breaking of the global $SU(3)$ symmetry, such that the Higgs 
triplets ${\cal H}_u$ and ${\cal H}_d$ acquire vacuum expectation values in 
the $S_u$ and $S_d$ directions 
\beq
f_u\equiv\langle S_u\rangle = f\sin\beta~,\phantom{aaa}
f_d\equiv\langle S_d\rangle = f\cos\beta~,\label{eqn:susdvevs}
\eeq
can be achieved in a similar way as the electroweak symmetry breaking in 
the MSSM, i.e. by quantum corrections induced by the large Yukawa coupling
$Y_T$ in the superpotential (\ref{eqn:sup_wt}) \cite{BICHGA}. Similarly as 
in the MSSM, the mass of the additional neutral $Z^\prime$ boson is then 
given by:
\begin{eqnarray}
{1\over2}M^2_{Z^\prime}=g_E^2f^2={m_1^2 - m_2^2\tan^2\beta\over\tan^2\beta-1}
={m_1^2 + m_2^2\over2} 
\left[ {m_1^2 - m_2^2\over\sqrt{ (m_1^2 + m_2^2)^2- 4 m_3^4}} -1 \right]
\nonumber\\
\sin2\beta = {2m_3^2\over m_1^2 + m_2^2} 
\phantom{aaaaaaaaaaaaaaaaaaaaaaa}
\end{eqnarray}

There is also the tree-level $SU(3)$ breaking 
potential originating from the electroweak $D$-terms:
\begin{eqnarray}
V_{EW} = {g^2_2 + g^2_y\over8}(|H_u|^2 -|H_d|^2)^2  
- {g^2_2\over2} (|H_d H_u|^2 - |H_u|^2|H_d|^2)~.  \label{eqn:V_D}
\end{eqnarray}

Let us now identify the SM Higgs doublet. The $SU(3)$ global symmetry is 
spontaneously broken down to $SU(2)$ by the vevs (\ref{eqn:susdvevs}). Hence 
five pseudo-Goldstone bosons emerge, of which one becomes the longitudinal 
component of the massive $U(1)_E$ gauge boson. The four remaining physical 
degrees of freedom are identified with the SM Higgs doublet. Below the scale 
$f$ of the $SU(3)$ breaking we can work with the non-linear realization of 
the $SU(3)$ symmetry and ignore all heavy scalars from the Higgs sector. In 
this approach the Higgs triplets are parametrized as
\begin{eqnarray}
{\cal H}_u = e^{i\Pi/f} \left(\matrix{0\cr f_u}\right),
\phantom{aaa}
{\cal H}_d = \left(0,~ f_d\right)e^{-i\Pi/f}
\end{eqnarray}
where\footnote{The scalar field associated with the broken generator $T^8$ 
of $SU(3)$ becomes the longitudinal component of the massive $Z^\prime$
boson and needs not be included in $\Pi$.}
\begin{eqnarray}
i\Pi = \left(\matrix{{\bf 0_{2\times2}} & H \cr
                     -H^\dagger & 0} \right) 
\end{eqnarray}
which leads to
\begin{eqnarray}
\left(\matrix{H_u\cr S_u}\right)=\sin\beta
\left(\matrix{H{\sin(|H|/f)\over |H|/f }\cr\phantom{a}\cr f\cos(|H|/f)}\right)
\phantom{aaa}
\left(\matrix{H_d\cr S_d}\right)=\cos\beta
\left(\matrix{H^\ast{\sin(|H|/f)\over|H|/f}\cr\phantom{a}\cr 
f\cos(|H|/f)}\right)
\end{eqnarray}
where $|H|\equiv\sqrt{|H^\dagger H|}$. In the following we will keep track 
only of the real neutral component of the Higgs doublet $H$, i.e. we will 
substitute $(0,h)^T$ for $H$ and $h$ for $|H|$. With this parametrization it 
is explicit that the $SU(3)$ preserving Higgs potential (\ref{eqn:V_su3symm}) 
does not contribute to the potential  of the SM Higgs doublet $H$. At 
the tree level the SM Higgs potential  has only the quartic part which 
arises from the electroweak $D$-terms (\ref{eqn:V_D}) explicitly breaking 
global $SU(3)$: 
\begin{eqnarray}
V_{\rm tree} ={1\over8}(g^2_2 + g^2_y)\cos^2(2\beta) f^4 \sin^4(h/f)~.
\label{e.tlqh}
\end{eqnarray}

\section{One loop SM Higgs potential}

Since the $SU(3)$ symmetry is only approximate, corrections to the SM 
Higgs potential appear at loop level. We therefore calculate the one-loop 
effective potential $V=V_{\rm tree}+\Delta V_{\rm1-loop}$ in terms of the 
Lagrangian parameters renormalized at the scale $\Lambda_S$. In a 
supersymmetric model such a calculation may be equivalently viewed as the 
calculation in terms of the bare parameters and with the momentum cut-off 
$\Lambda_S$. In a consistent one-loop calculation of the effective potential  
the $SU(3)$ symmetric parametres defined by eq. (\ref{eqn:softmasses}), 
(\ref{eqn:V_su3symm}), (\ref{eqn:sup_mu}), (\ref{eqn:sup_wt}) and 
(\ref{eqn:sup_wb}), must be used. An $SU(3)$ splitting of these parameters 
is generated at one loop level, too, by the two sources of explicit $SU(3)$ 
breaking: the nonzero electroweak gauge couplings $g_2$, $g_y$ and the 
$\mu_T$ and $\mu_B$ terms. It enters the effective potential only in the 
two-loop approximation. For the mechanism of the electroweak symmetry 
breaking we propose in this paper such higher order effects must be 
negligible and this constrains the scale $\Lambda_S$ at which the soft mass 
terms are generated. As we shall see, for squarks masses ${\cal O}(10$~TeV) 
our mechanism works anyway only for $\Lambda_S\sim{\cal O}(100$~TeV),
consistently with the above requirement. Moreover, we assume that the $SU(3)$ 
breaking corrections to the Yukawa couplings also vanish above $\Lambda_S$, 
i.e. that above $\Lambda_S$ the model described by eqs. (\ref{eqn:sup_wt}), 
(\ref{eqn:sup_wb}), (\ref{eqn:softmasses}), (\ref{eqn:sup_mu}) and 
(\ref{eqn:V_su3symm}) is embedded in some  $SU(3)$ invariant theory as in 
\cite{BICHGA}.

Before presenting our results for the one-loop effective potential we 
recall the structure of the effective potential in other models. In the MSSM
quadratically divergent corrections to the Higgs mass parameter are absent
at any order of perturbation theory due to supersymmetry. Logarithmically 
divergent contribution is determined by ${\rm STr}{\cal M}^4$ and depends 
quadratically on the supersymmetry breaking mass parameters. It consists 
of two parts: one proportional to the top Yukawa coupling quadratically 
dependent on the sfermion and Higgs fields soft mass parameters and one 
proportional to the  gauge couplings and quadratically dependent on gaugino 
masses. In the non-supersymmetric Little Higgs Models \cite{arcoge} 
global symmetries forbid quadratically divergent corrections to 
the Higgs mass parameter at one-loop (but such divergences are present 
already at two-loops). In the language of the effective potential, the mass 
matrix squared ${\cal M}^2$ in these models does not depend on the SM Higgs 
field $h$ and, what is important, the cancellations occur independently in 
fermionic and bosonic sectors. Logarithmically divergent corrections 
proportional to the top Yukawa coupling are generically present already at 
the one-loop level and they, as well as the two-loop quadratically divergent 
ones, require some ultraviolet completion of the models at the scales of 
order 10~TeV. Finally, there are models with ``hard'' supersymmetry breaking 
but such, that the leading contribution to the effective potential 
proportional to the top quark Yukawa coupling is finite at one-loop 
\cite{bahano}. However quadratically divergent contributions appear from 
$D$-terms \cite{GHNINI} and at higher orders and require a low cut-off.

In the model discussed in this paper the situation is still different 
and more elegant, as long as all $SU(3)$ breaking quantum effects in the 
parameters entering $\Delta V_{\rm1-loop}$ can be neglected (i.e. for a 
low scale $\Lambda_S$). Because the model is supersymmetric, all quadratic 
divergences are absent to all orders in perturbation theory. Moreover, the 
SM Higgs potential must be proportional to some parameter that breaks $SU(3)$ 
symmetry and also leads to supersymmetry breaking for $h\neq0$ (in the sense 
of forcing some $F$- or $D$-terms to be nonvanishing for $h\neq0$). The most 
important consequence is that the contribution proportional to the top (and 
also bottom) Yukawa coupling is finite at one-loop and only logarithimically 
sensitive to the sfermion mass scale $m_{\tilde q}^2$. Indeed, since the 
$Y_T$ Yukawa coupling and stop soft mass terms are $SU(3)$ symmetric the Higgs 
potential should be proportional to $Y_T^2\mu_T^2$. But $\mu_T$ does not lead 
to supersymmetry breaking for $h\neq0$, therefore such a contribution cannot 
occur in ${\rm STr}{\cal M}^4$. It can only enter the finite part of the Higgs 
potential such as $\Delta V_{\rm 1-loop}\sim Y_T^2\mu_T^2h^2 
\log(m_{\tilde q}/f)$. The mild logarithmic sensitivity to $m_{\tilde q}$ 
allows us to raise the squark masses far above 1~TeV without introducing too 
much fine-tuning. Logarithmically divergent contributions that arise at one 
loop are suppressed by electroweak gauge couplings. In our model with a 
low cutoff scale $\Lambda_S$, these will be of the same order of magnitude 
as the finite part dependent on the top Yukawa couplings.  

Let us first discuss the top/stop contribution to the one-loop correction 
$\Delta V_{\rm1-loop}$ to the SM Higgs potential in our model. As usually, 
it is given by:
\begin{eqnarray}
\Delta V_{\rm1-loop} = {1\over64\pi^2}{\rm STr}\left\{{\cal M}^4 
\left(\log{{\cal M}^2\over\Lambda^2_S}-{3\over2}\right)\right\}  ~.
\label{eqn:V1loop}
\end{eqnarray}
where ${\cal M}^2$ is the field dependent mass squared matrix of the theory.
For a nonzero background value $h$ of the SM Higgs field the contribution
of the top sector to the fermionic mass matrix reads
\begin{eqnarray}
\cl_{\rm mass}=-(t^c, T^c) \cm_{\rm tops} (t,T)^T +\hc \phantom{aaa}
\cm_{\rm tops} = \left(\matrix{Y_T f_u s_h  &  Y_T f_u c_h \cr
                                0       &            \mu_T}\right)
\end{eqnarray}
where we have used te abbreviations $s_h=\sin(h/f)$, $c_h=\cos(h/f)$.
Hence,
\begin{eqnarray}
\cm_{\rm tops}^2\equiv\cm^\dagger_{\rm tops}\cm_{\rm tops} =
\left(\matrix{Y_T^2 f_u^2 s_h^2    &  Y_T^2 f_u^2 s_h c_h \cr
              Y_T^2 f_u^2 s_h c_h  &  \mu_T^2 + Y_T^2 f_u^2 c_h^2}\right)~.
\label{eqn:topMM}
\end{eqnarray}
In the stop sector
\begin{eqnarray}
\cl_{\rm mass} = - (\tilde t^\ast,\tilde T^\ast, \tilde t^c, \tilde T^c) 
\cm^2_{\rm stops} (\tilde t,\tilde T,\tilde t^{c\ast},\tilde T^{c\ast})^T  
\end{eqnarray}
where for vanishing gauge couplings $\cm^2_{\rm stops}$ takes the form
\begin{eqnarray}
\left(\matrix{
m_{{\cal Q}_3}^2 + Y_T^2 f_u^2 s^2_h &  Y_T^2 f_u^2 s_h c_h  & 
(Y_T f_d \mu - A_T f_u) s_h & 0
\cr
Y_T^2 f_u^2 s_h c_h  &  m_{{\cal Q}_3}^2 + \mu_T^2 + Y_T^2 f_u^2 c_h^2 & 
(Y_T f_d \mu  - A_T f_u) c_h & 0 
\cr
(Y_T f_d \mu - A_T f_u) s_h & (Y_T f_d \mu - A_T f_u) c_h & 
m_{U_3}^2+ Y_T^2 f_u^2  & Y_T f_u \mu_T c_h
\cr
0 & 0 & Y_T f_u \mu_T c_h & m_{T^c}^2 + \mu_T^2}\right)\nonumber\\
\label{eqn:stopM2}
\end{eqnarray}

Let us denote the two eigenvalues of the ($h$ dependent) top mass matrix 
squared (\ref{eqn:topMM}) by $t_1^2$ and $t_2^2$ and the four eigenvalues of 
the stop mass squared matrix (\ref{eqn:stopM2}) by $ m^2_{\tilde q}+s^2_i$, 
where $m_{\tilde q}$ is the overall scale of the soft supersymmetry breaking 
in the stop sector. In the following we assume\footnote{The reasoning used 
here implicitly assumes that the soft mass terms in the stop sector are 
almost degenerate, but as long as the entries mixing left and and right stops 
in the matrix (\ref{eqn:stopM2}) are small, the results are independent of 
this assumption. This can be checked by directly computing the eigenvalues of 
(\ref{eqn:stopM2}).}
that $m^2_{\tilde q}\gg s^2_i$ and expand the one-loop effective potential 
in powers of $1/ m^2_{\tilde q}$. Up to terms of order $1/m^2_{\tilde q}$ we 
can rewrite the top/stop sector contribution as:
\begin{eqnarray}
\Delta V_{\rm 1-loop} = {N_c \over 32\pi^2}\left\{
\left[{\rm Tr} \cm_{\rm stops}^4 - 2{\rm Tr} \cm_{\rm tops}^4 \right]
\log{m^2_{\tilde q}\over\Lambda_S^2}
- 2 \sum_{i=1}^2 t_i^4\left[\log{t_i^2\over m^2_{\tilde q}}-{3\over2}
\right]\right\}  \nonumber\\
+{\rm const.} + \co(1/m^2_{\tilde q})  ~.\phantom{aaaaaaaaaaaaaaaaaaaa}
\label{eqn:V1_loop_expanded}
\end{eqnarray}
where the color factor $N_c=3$. We also used the fact that both 
${\rm Tr}\cm_{\rm stops}^2$ and 
${\rm Tr}\cm_{\rm tops}^2$ do not depend on $h$ as a consequence of $SU(3)$ 
symmetry. As explained before, Tr$\cm_{\rm stops}^4-2{\rm Tr}\cm_{\rm tops}^4$ 
does not depend on $Y_T$. Therefore, in order to calculate the part of 
$\Delta V_{\rm 1-loop}$ proportional to the Yukawa couplings in the limit 
$m^2_{\tilde q}\gg s^2_i$ it is sufficient to find the eigenvalues of 
the top mass matrix squared (\ref{eqn:topMM}).
  
The necessary eigenvalues of the top  mass matrix are given by:
\beq
t_{1,2}^2 = {1\over2}\left[Y_T^2 f_u^2 + \mu_T^2 
\pm\sqrt{ (Y_T^2 f_u^2 + \mu_T^2)^2 - 4 \mu_T^2 Y_T^2 f_u^2 s^2_h } \right] 
\label{e.te}
\eeq
The lower sign corresponds to the ordinary top quark mass, which 
approximately equals $m_t \approx y_t f \sin(h/f)$
where  
\beq
y_t = {\mu_T Y_T \sin\beta \over\sqrt{Y_T^2 f_u^2 + \mu_T^2}}
\label{e.tse}
\eeq
is the physical top quark Yukawa coupling. The higher eigenvalue 
\beq
\label{e.mT}
t_2^2\approx m_T^2\equiv Y_T^2 f_u^2 + \mu_T^2\label{eqn:mTdef}
\eeq
corresponds to the mass squared of the $SU(3)$ fermionic partner of the 
top quark. Inserting (\ref{e.tse}) in the last term in the curly brackets 
in (\ref{eqn:V1_loop_expanded}) we can determine the contribution to the 
effective potential proportional to the top Yukawa coupling $Y_T$. 
Expanding in powers of $\sin(h/f)$ to the quartic order we find:  
\begin{eqnarray}
\Delta V_{\rm1-loop}^{(1)} = 
-{3\over8\pi^2} Y_T^2 \mu^2_T f^2 \sin^2\beta
\left(\log {m_{\tilde q}^2\over m^2_T} +1\right) \sin^2(h/f)\phantom{aaa}
\nonumber\\
-{3\over8\pi^2}{Y^4_T\mu^4_T\sin^4\beta \over m_T^4} 
f^4\left[\log\left({Y_T\mu_Tf\sin\beta\over m_T^2}\sin(h/f)\right) 
+{1\over4}\right] \sin^4(h/f)
\label{e.hpyc}
\end{eqnarray}
The first term of (\ref{e.hpyc}) gives the $Y_T$ dependent one-loop 
contribution to the SM Higgs mass squared $m_H^2$. It is negative, which 
enables the electroweak symmetry breaking. Moreover, as advertised, it 
is not proportional to $m_{\tilde q}^2$ but rather to the square of the 
$SU(3)$ breaking supersymmetric parameter $\mu_T$, which may be much smaller 
than the soft supersymmetry breaking scale $m_{\tilde q}$. The analogous 
contribution from the bottom-sbottom sector, can be easily obtained by 
using the substitutions: $Y_T\to Y_B$, $\mu_T\to\mu_B$, $f_u\to f_d$, 
$\sin\beta\to\cos\beta$. It is however small compared to the 
top-stop sector because $Y_B\cos\beta\ll Y_T\sin\beta$.

Terms in $\Delta V_{\rm 1-loop}$ logarithmically depending on the scale 
$\Lambda_S$, i.e. proportional to STr$\cm^4$ arise from two different 
sources which depend on the $SU(3)$ breaking SM gauge interactions. One is 
the gauge coupling dependent contribution to the sfermion mass matrices, 
that enters via ${\rm Tr} \cm_{\rm stops}^4 -2 {\rm Tr} \cm_{\rm tops}^4$ in 
(\ref{eqn:V1_loop_expanded}). In this way we get the $h$ dependent contribution
\begin{eqnarray}
\Delta V_{\rm1-loop}^{(2)} = 
{1\over32\pi^2} g^2_y \cos(2\beta) f^2 {\rm Tr}[Y m^2] 
\log{\Lambda_S^2\over m_{\tilde q}^2} \sin^2(h/f) \nonumber\\
-{3\over64\pi^2}(g_2^2+g_y^2)\cos(2\beta)\sin^2\beta f^4 Y^2_T
\log{\Lambda_S^2\over m_{\tilde q}^2} \sin^4(h/f)
\label{e.hpdc}
\end{eqnarray}
where the trace runs over all sfermions charged under $U(1)_Y$. We have 
dropped the terms with fourth powers of the gauge couplings as they are 
small (potentially large terms $\propto g_2^2g_E^2$ cancel out between 
the up and down type squark contributions). For non-universal soft breaking 
scalar masses, when Tr$[Y m^2]\neq0$, this term depends quadratically 
on the scale $m_{\tilde q}$. However it is suppressed by the small coupling 
$g_y$. It is also interesting to notice that in contrast to the Little
Higgs models, in which a complicated gauge structure is used to cancel the
$g_2^2\Lambda^2$ contribution to $m_H^2$, here it is the supersymmetric 
structure alone which ensures the absence of the $g_2^2m^2_{\tilde q}$
piece.

The complete $h$ dependent contribution of the gaugino/gauge boson and 
higgsino/Higgs boson sectors to $\Delta V_{\rm1-loop}$ is rather 
lengthy.\footnote{In order to compute the Higgs sector contribution to 
$\Delta V_{\rm1-loop}$, taking into account also heavy degrees of freedom 
necessary for vanishing of the $h$ dependent part of STr$\cm^2$, we split 
the neutral CP-even components of the fields in (\ref{eqn:V_su3symm}) and 
(\ref{eqn:V_D}) into $H^0_{u,d}+$quantum fluctuations, $S_{u,d}+$quantum 
fluctuations, compute the mass squared matrices of the quantum neutral 
CP-even and CP-odd and charged Higgs fields and only at the end substitute 
$H^0_{u,d}\to f_{u,d}s_h$, $S_{u,d}\to f_{u,d}c_h$.}
Its most relevant parts can be approximated by
\begin{eqnarray}
\Delta V_{\rm1-loop}^{(3)}={1\over64\pi^2}
\left[3(3g^2+g_y^2)\mu^2 + 12 g_2^2 M_2^2+4g_y^2M_y^2 \right.
\phantom{aaaaaaa}\nonumber\\
-4 \sin2\beta(3g_2^2M_2+g_y^2M_y)~\mu\phantom{aaaaaaaaaaaa}\nonumber\\
-g^2_E(3g_2^2+4g_y^2)\cos^22\beta f^2
\left.\right] f^2
\log{\Lambda_S^2\over\mu^2} \sin^2(h/f) \label{e.hpgc}\\
-{1\over64\pi^2}g^2_E(g_2^2+g_y^2) f^4
\log{\Lambda_S^2\over\mu^2} \sin^4(h/f)\phantom{aaaaaaaa}\nonumber
\end{eqnarray}
We have neglected the contributions proportional to the soft mass terms of the
Higgs fields and those with electroweak gauge couplings $g_2$ and $g_y$ in 
the fourth power. We have also approximated $\cm^2$ under the logarithm in 
(\ref{eqn:V1loop}) by $\mu^2$.

\section{$SU(3)$ and electroweak breaking}
\label{sec:4}

We turn to analizing in a more quantitive way radiatively induced $SU(3)$ and 
electroweak breaking. Our basic assumption is that the sfermion soft masses 
set the largest mass scale in the model. Breaking of the gauge 
$U(1)_E$ and the global $SU(3)$ symmetries occurs if $m_1^2 m_2^2 <m_3^4$ in 
eq. (\ref{eqn:V_su3symm}). This condition is obviously satisfied for $m_1^2 >0$
and $m_2^2<0$. Similarly as in the MSSM, the quantum corrections generated 
primarily by the top/stop loops can make the parameter $m^2_{{\cal H}_u}$ 
negative. In the present model this effect will be enhanced by the largeness 
of soft squark masses. The leading one-loop effect of the renormalization of  
$m^2_{{\cal H}_u}$ between the scales $\Lambda_S$ and $m_{\tilde q}$ is 
given by 
\begin{eqnarray}
\label{e.mhurun}
m^2_{{\cal H}_u}(m_{\tilde q})\approx m^2_{{\cal H}_u}(\Lambda_S)
-{1\over(4\pi)^2}\left[g_E^2{\rm Tr}(Em^2) + 6Y_T^2(m_{\cal Q}^2 + m^2_{U_3})
\right] \log {\Lambda_S \over m_{\tilde q}}  
\end {eqnarray}
We have not displayed here the much smaller $g_y^2{\rm Tr}(Y m^2)$ and the
gaugino contribution. Moreover, since we will assume that 
$m^2_{{\cal H}_u}(\Lambda_S)\ll m_{\tilde q}^2(\Lambda_S)$, we have dropped 
the Higgs soft masses in the second term of \eref{mhurun} as well. With this 
assumption, as long as ${\rm Tr}(E m^2)$ is not too large and negative, the 
soft mass $m_{{\cal H}_u}^2$ is driven towards a  negative value. For very 
large $\tan \beta \sim 50$ one should check whether $m_{{\cal H}_d}^2$ is not 
driven to negative values too but even in that case the situation can be 
improved for positive ${\rm Tr}(Em^2)$. 

For $\tan\beta\geq5$ and $Y_T^2\sim 1.3$ (see 
below) the scale $f$ of the $SU(3)$ breaking can be estimated as:
\beq
\label{e.f}
g_E^2f^2 \approx - m_2^2 \approx -\mu^2 + 
0.1 m_{\tilde q}^2 \log{\Lambda_S\over m_{\tilde q}} 
\eeq  
Since the $U(1)_E$ gauge coupling $g_E$ does not contribute to the SM Higgs 
mass it does not have to be small and in the following we assume 
$g_E\approx1$. This choice minimizes the fine-tuning necesary for $SU(3)$ 
breaking. Moreover, we shall soon see that increasing $f$ increases 
fine-tuning in the electroweak symmetry breaking. Therefore adopt 
the minimal phenomenologically allowed value $f\sim2.5$~TeV \cite{APDOHO}. 
Then \eref{f} sets the lower bound on  the squark masses 
$m_{\tilde q}>8/\log^{1/2}(\Lambda_S/m_{\tilde q})$~TeV. It also correlates 
the values of $\mu$, $m_{\tilde q}$ and $\Lambda_S$ so that 
$\mu^2\sim0.1 m_{\tilde q}^2\log ({\Lambda_S/  m_{\tilde q}})$. 
Essentially no fine-tuning is required for $f\approx2.5$~TeV as long as 
$m_{\tilde q}\simlt10$~TeV and $\Lambda_S \sim 100 \tev$. A fine-tuning of 
order $1\%$ is required for $m_{\tilde q}\sim50$~TeV.

In the second stage we study breaking of the electroweak symmetry. The 
leading contributions to the SM Higgs potential are those of \eref{hpyc}, 
\eref{hpdc} and \eref{hpgc}. The resulting SM Higgs doublet potential
has the approximate form: 
\beq
V = m^2_H h^2 + \lambda h^4 + \kappa h^4 \log{h^2\over m_T^2}
\eeq
At tree-level $m^2_H=0$, $\lambda={1\over8}(g^2_2+g_y^2)\cos^2(2\beta)$ 
and $\kappa=0$. At one loop $\Delta V_{\rm1-loop}^{(1)}$, 
$\Delta V_{\rm1-loop}^{(2)}$ and $\Delta V_{\rm1-loop}^{(3)}$ contribute 
to $m^2_H$: 
\begin{eqnarray}
\label{e.m2}
m^2_H \approx -{3\over4\pi^2} y_t^2 m_T^2 \log{m_{\tilde q}\over m_T} 
+{g^2_y\over16\pi^2}\cos(2\beta){\rm Tr}[Y m^2]
\log{\Lambda_S\over m_{\tilde q}}\nonumber\\
+{1\over32\pi^2}\left[3(3g_2^2+g_y^2)\mu^2 +12 g_2^2 M_2^2+4g_y^2M_y^2\right.
\phantom{aaaaa}\nonumber\\
-4 \sin2\beta(3g_2^2M_2+g_y^2M_y)~\mu\phantom{aaaaaaaaa}\nonumber\\
\left.-g^2_E(3g_2^2+4g_y^2)\cos^22\beta f^2\right]\log{\Lambda_S\over\mu}
\phantom{aa}
\end{eqnarray}
In \eref{m2} the Higgs mass parameter at the electroweak scale is expressed 
in terms of the other parameters renormalized at the scale $\Lambda_S$. In 
the first term we have used the tree-level relations \erefn{tse} and 
(\ref{eqn:mTdef}) to express $Y_T(\Lambda_S)$ in terms of the top quark 
Yukawa coupling $y_t(m_t)$, consistently with the systematic  one-loop 
calculation of $\Delta V_{\rm 1-loop}$. Then for $m_t=(178\pm4.3)$~GeV
\cite{tev} and the corresponding $\overline{\rm MS}$ mass $169$~GeV we have
\beq
\label{e.yt}
{\mu_T Y_T\sin\beta \over \sqrt{Y_T^2 f_u^2 + \mu_T^2}}=0.97
\eeq 
which relates the values of $Y_T(\Lambda_S)$ and $\mu_T(\Lambda_S)$. We can 
now analyze the contribution of the first term on the rhs of \eref{m2}. It 
is quadratically dependent on the mass $m_T$ of the heavy top quark partner. 
For $f=2.5$~TeV, taking into account the relation \erefn{yt}, $m_T$ has a 
minimum for $\mu_T\approx3$~TeV and $Y_T\approx1.35$ at which the 
contribution (\ref{e.hpyc}) to the Higgs mass parameter is 
$m_H^2\approx-1.6\tev^2$. We choose this value of $\mu_T$ as it minimizes the 
fine-tuning in electroweak breaking. Note that these arguments do not depend 
on the value of stop masses $m_{\tilde q}$. Also perturbativity of $Y_T$, i.e. 
$Y_T^2/4\pi<1$ up to the scale $\Lambda_S$, does not impose any new bound.

Since for the electroweak breaking we need $m_H^2\sim0.01\tev^2$ the finite 
Yukawa contribution is by itself too large by a factor of hundred. For our 
mechanism to work, it must be partly cancelled by the other contributions in 
\eref{m2}. Thus, our first conclusion is that fine-tuning at $1\%$ level is 
required here. However the squark masses can be large. Note that in the MSSM 
for squark masses $m_{\t q}\sim10$~TeV the fine-tuning is of order $0.01\%$.

Before discussing the remaining terms in \eref{m2} it is important to note 
that the scale dependence of the top quark Yukawa coupling is relatively 
strong. Although this is formally a two-loop effect, it may introduce 
important corrections  to our systematic one-loop calculation. To estimate 
this uncertainty we may use \eref{tse} for $y_t(\Lambda_S)$ rather than 
$y_t(m_t)$, where $y_t(\Lambda_S)$ is obtained by evolving the top Yukawa 
coupling from the scale $m_t$ to $\Lambda_S$ using the appropriate RG 
equations. Then the corresponding numbers are $\mu_T\approx3$~TeV, 
$Y_T\approx1.15$ and the contribution (\ref{e.hpyc}) to the mass parameter 
is $m_H^2\approx-1.2\tev^2$. The fine-tuning is then slightly smaller 
but still of order 1\%. Of course, using the present procedure for 
extracting $y_T(\Lambda_S)$ from the physical top quark mass we should, for 
consistency, calculate the effective potential at two loops. However, since 
the loop corrections to the top Yukawa coupling are likely the most important, 
the comparison of the two methods is at least a good estimate of the 
uncertainty of our results.  
      
Turning now to the other contributions in \eref{m2}, the second one in 
importance is the $\mu^2$ dependent part of the last term. Indeed, because 
\eref{f} fixes $\mu^2$ as a function of $m_{\t q}$ and $\Lambda_S$, this 
contribution reads approximately $10^{-3}m_{\t q}^2\log^2(\Lambda_S/ m_{\t q})$
and for heavy sfermions  it is similar in magnitude (but with opposite sign)  
to the Yukawa contribution. In fact, the requirement that it does not make 
$m_H^2$ positive sets a stringent bound on the cutoff scale $\Lambda_S$, 
$\Lambda_S\simlt m_{\t q} e^{33\tev/m_{\t q}}$. For $m_{\tilde q}=10$~TeV, 
we get $\Lambda_S\simlt250$~TeV. This is a post-factum justification of 
our assumption that the cutoff scale is not high. As for the remaining 
contributions in \eref{m2}, the term proportional to ${\rm Tr}[Ym^2]$ is 
small even for non-universal squark masses. In that case we would expect 
${\rm Tr}[Ym^2]\sim m_{\t q}^2$, but it is strongly suppressed by the loop 
factor and the small hypercharge gauge coupling. The positive 
contribution of gaugino masses to $m_H^2$ can be larger and can complement 
the $\mu^2$ contribution in canceling the too large negative contribution 
of (\ref{e.hpyc}). For $\Lambda_S\approx100$~TeV the requirement that 
$m_H^2<0$ puts the upper bound $M_2\simlt4.5$~TeV.

Collecting all the relevant contributions to $m_H^2$ we see that with a $1\%$
fine-tuning the necessary value for this parameter $m_H^2\sim(100$~GeV$)^2$
is easy to obtain for $m_{\tilde q}\sim{\cal O}(10$~TeV) and 
$\Lambda_S\sim{\cal O}(100$~TeV).

For the one-loop contributions to the quartic couplings $\lambda$ and $\kappa$
we get\footnote{The corrections to the quartic coupling $\lambda$ contained in 
the $\sin^2(h/f)$ terms in \eref{hpyc}, \eref{hpdc} and \eref{hpgc}, once the 
large contribution to $m_H^2$ proportional to $Y_T^2\mu_T^2$ is canceled 
against other contributions, becomes of order 
$m_H^2/6f^2\sim-(100 ~{\rm GeV})^2/6(2.5~{\rm TeV})^2$ and is negligible.}
\begin{eqnarray}
\label{e.lk}
\lambda={1\over8}(g_y^2+g_2^2)\cos^22\beta-{3\over8\pi^2} y_t^4  
\left[\log\left(y_t \right) 
+{1\over4}\right]\phantom{aaaa}\nonumber\\ 
-{3\over64\pi^2}(g_2^2+g_y^2)Y^2_T\cos(2\beta)\sin^2\beta 
\log{\Lambda_S^2\over m_{\tilde q}^2}  
-{1\over64\pi^2}g^2_E(g_2^2+g_y^2) 
\log{\Lambda_S^2\over\mu^2}\nonumber\\
\kappa=-{3\over16\pi^2} y_t^4
\phantom{aaaaaaaaaaaaaaaaaaaaaaaaaaaaa}
\end{eqnarray}
From this we can estimate the SM Higgs  mass:
\begin{eqnarray}
M_h^2 = 2 v^2 \left(\lambda + {3\over2}\kappa +  
\kappa\log(v^2/2 m_T^2)\right)
\end{eqnarray}
The term proportional to $\kappa\log(v^2/2 m_T^2)$ turns out to be the 
dominant correction to 
the SM Higgs mass. Its effect is to change the tree-level prediction for the 
Higgs mass, and  to raise it above $M_Z$. Note that $\kappa$ in \eref{lk} is 
set by the top quark Yukawa coupling $y_t$ and so this contribution depends 
only logarithmically on $m_T$ (and thus on $\mu_T$ and $f$).  Inserting the 
numbers we get the Higgs mass (for $\tan\beta\geq5$):
\beq
M_h \approx 120 - 135  \gev \, . 
\eeq
This is the prediction of present model. The lower value is obtained when the 
higher order effects associated with the RG running of the top quark Yukawa 
copling from $m_t$ to $\Lambda_S$ is taken into account. For this reason we
believe it approximates slightly better the true Higgs mass in our model.
There is of course also the dependence of $M_h$ on 
$\tan\beta$, which becomes significant for $\tan\beta\leq5$. 

\section{Conclusions} 

In this paper we have discussed  electroweak symmetry breaking in a 
supersymmetric 
model  in which the SM Higgs doublet is a pseudo-Goldstone boson of $SU(3)$ 
global symmetry. The Higgs mass  parameter is generated at one loop level by 
two different, moderately fine-tuned sources of the global symmetry breaking. 
The mechanism works well for heavy sfermion masses $m_{\t q}\sim 10$~TeV, but 
the fine-tuning is, nevertheless, of order 1\%, two orders of magnitude less 
than in the MSSM with similar sfermion masses. The scale $\Lambda_S$ at which 
supersymmetry breaking is mediated to the visible sector has to be low, of 
order $100$~TeV.

Several of the phenomenological consequences of our scenario are similar 
to those of the so-called split supersymmetry model of Arkani-Hamed and 
Dimopoulos \cite{ardi}. The sfermions as well as additional scalars in the 
Higgs sector should be beyond reach of the LHC. Also, the heavy top quark 
partner has mass $m_T\simgt4$~TeV, too large to be seen in the LHC 
\cite{atlas}. Chargino and neutralino masses are more model dependent but 
they should be smaller than 4~TeV. The important difference with the split 
supersymmetry scenario is that although the gluino mass is not bounded by
the mechanism of the electroweak symmetry breaking, gluino is not expected
to be long lived because squarks are not so heavy. The Higgs boson mass is 
predicted (for $\tan\beta\gg1$) in the range $120-135$~GeV, whereas it is 
in the range $130-170\gev$ in split supersymmetry \cite{giro}. Another 
potential signature of our mechanism is the presence of the $Z^\prime$ gauge 
boson with $m_{Z^\prime}\sim3$~TeV, which should easily be discovered in the 
LHC 
\cite{atlas}.     
   
\section*{Acknowledgments}

S.P. would like to thank the Physics Departments at the University of Bonn
and at the University of Hamburg and the APC Institute at the University 
Paris VII for their hospitality. His visits to the University of Bonn
and at the University of Hamburg were possible thanks to the research 
award of the Humboldt Foundation.  
P.H.Ch. was partially supported by the RTN European Program HPRN-CT-2000-00152 
and by the Polish KBN grant 2 P03B 040 24 for years 2003--2005. A.F. and S.P. 
were partially supported by the RTN European Program HPRN-CT-2000-00148 and by 
the Polish KBN grant 2 P03B 129 24 for years 2003--2005.

\end{document}